\documentclass[authoryear,12pt]{elsarticle}

\usepackage{graphicx,footmisc}
\usepackage{subfig,epstopdf,multiequations}
\usepackage{natbib,url}
\usepackage{lineno,amsmath,amssymb}
\title{Inertia-less convectively-driven dynamo models in the limit of low Rossby number and large Prandtl number}%

\author[1]{Michael A. Calkins \fnref{e1} \corref{c1}}
\fntext[e1]{email:  michael.calkins@colorado.edu}
\author[2]{Keith Julien \fnref{e2}}
\fntext[e2]{email:  keith.julien@colorado.edu}
\author[3]{Steven M. Tobias \fnref{e3}}
\fntext[e3]{email:  s.m.tobias@leeds.ac.uk}

\cortext[c1]{Corresponding author}

\address[1]{Department of Physics, University of Colorado, Boulder, CO 80309 USA}
\address[2]{Department of Applied Mathematics, University of Colorado, Boulder, CO 80309 USA}
\address[3]{Department of Applied Mathematics, University of Leeds, Leeds, UK LS2 9JT}

\journal{Phys.~Earth Planet.~Int.}

\newcommand*{\be}{\begin{equation}}
\newcommand*{\ee}{\end{equation}}

\def\begineq{\begin{equation}}
\def\endeq{\end{equation}}

\def\begineqn{\begin{equation*}}
\def\endeqn{\end{equation*}}

\def\beginar{\begin{eqnarray}}
\def\endar{\end{eqnarray}}
\def\beginarn{\begin{eqnarray*}}
\def\endarn{\end{eqnarray*}}

\def\lb{\left ( }
\def\rb{\right ) }

\def\ep{\epsilon}

\def\Rat{\widetilde{Ra}}

\def\qt{\widetilde{q}}

\def\ub{\mathbf{u}}

\def\Bb{\mathbf{B}}
\def\ubp{\mathbf{u}^{\prime}}
\def\bbp{\mathbf{b}^{\prime}}
\def\thp{\theta^{\prime}}

\def\rot{\mathbf{\Omega}}

\def\mBb{\overline{\bf B}}

\def\mP{\overline{p}}

\def\mth{\overline{\vartheta}}
\def\pth{\vartheta^{\prime}}
\def\pP{p^{\prime}}

\def\pfw{w^{\prime}}

\def\dt{{\partial_{T}}}
\def\dtau{{\partial_{\tau}}}
\def\dta{{\partial_\tau}}

\def\dsx{{\partial_x}}
\def\dsy{{\partial_y}}

\def\dst{{\partial_t}}

\def\dsz{{\partial_z}}

\def\dz{{\partial_Z}}
\def\dzt{{\partial^2_Z}}

\def\hz{{\bf\widehat z}}

\def\litx{{\bf x}}

\def\oAt{\frac{1}{A_\tau}}

\def\div{{\nabla \cdot}}

\def\lp{{\nabla_\perp^2}}

\def\wq{\widetilde{q}}
\def\mU{\mathcal{U}}

\begin{document}


\begin{abstract}

Compositional convection is thought to be an important energy source for magnetic field generation within planetary interiors. The Prandtl number, $Pr$, characterizing compositional convection is significantly larger than unity, suggesting that the inertial force may not be important on the small scales of convection as long as the buoyancy force is not too strong. We develop asymptotic dynamo models for the case of small Rossby number and large Prandtl number in which inertia is absent on the convective scale. The relevant diffusivity parameter for this limit is the compositional Roberts number, $q = D/\eta$, which is the ratio of compositional and magnetic diffusivities. Dynamo models are developed for both order one $q$ and the more geophysically relevant low $q$ limit. For both cases the ratio of magnetic to kinetic energy densities, $M$, is asymptotically large and reflects the fact that Alfv\'en waves have been filtered from the dynamics. Along with previous investigations of asymptotic dynamo models for $Pr=O(1)$, our results show that the ratio $M$ is not a useful indicator of dominant force balances in the momentum equation since many different asymptotic limits of $M$ can be obtained without changing the leading order geostrophic balance. Furthermore, the present models show that inertia is not a requirement for driving low $q$, large-scale dynamos.


\end{abstract}

\maketitle


\section{Introduction}

Planetary dynamos are thought to be generated by buoyancy-driven turbulence. Both thermal and compositional heterogeneities are possible buoyancy sources. The physical characteristics of thermal and compositional scalars are distinguished by their respective diffusion coefficients. Studies suggest that the thermal diffusivity in the Earth's outer core is larger than the compositional diffusivity by a factor of a thousand \citep{mP13}; i.e.~in non-dimensional terms, the Lewis number, $Le = \kappa/D = O(10^3)$, where $\kappa$ and $D$ are the thermal and compositional diffusivities, respectively. Denoting the fluid kinematic viscosity as $\nu$, the Prandtl numbers for the two scalars are therefore related via $Pr_C = Le Pr_T$, where $Pr_C = \nu/D = O(100)$ and $Pr_T = \nu / \kappa = O(10^{-1})$. 

Compositional convection has long been thought to play an important role in powering the geodynamo \citep{sB63,dS83,bB96b}. Recent studies have found that the thermal conductivity of the Earth's core may be larger than previously thought \citep{mP12,dK12}, suggesting that thermal convection may not play an important role in powering the geodynamo due to an associated increase in the conductive, adiabatic heat flux \citep{cD15}. Compositional convection has been suggested as the alternative power source for sustaining the geodynamo over the course of the Earth's evolution; this has stimulated interest in understanding the origin of such compositional heterogeneities \citep[e.g.][]{jO16}. There is clearly a need to better understand compositional convection and its influence on magnetic field generation. 

Provided the buoyancy forcing is not too strong, it is well known that the influence of inertia may be weak in large Prandtl number (thermal or compositional) convection \citep[e.g.][]{bS06}. The so-called large Prandtl number asymptotic limit can be used to remove inertia from the momentum equation; the resulting reduced system of equations are routinely employed for numerical investigations of convection in the Earth's subsolidus mantle where Prandtl numbers in excess of $O(10^{20})$ are typical \citep[e.g.][]{gS01}. In liquid metal planetary interiors the large Prandtl number limit is certainly invalid for the case of thermal convection, though it may be a useful approximation for describing compositional convection. We stress here, however, that the magnetic Prandtl number $Pm = \nu / \eta = O(10^{-5})$ in the Earth's core, where $\eta$ is the magnetic diffusivity. The large Prandtl number limit provides an interesting end-member case of convection in which all wave motion (inertial and magnetic) is absent; this approximation may then allow for assessing the influence that different waves might have on the dynamo when compared to cases that include inertia.

Many previous dynamo investigations have neglected the influence of inertia \citep[e.g.][]{kZ90,gG95b,cJ00b,jR02,dH16}. One of the arguments given in the literature for this simplification is that such effects are likely small in the Earth's core because the Rossby number, $Ro = \mU/2\Omega L$, is small ($\mU$ is a characteristic flow speed, $\Omega$ is the rotation rate, and $L$ is a characteristic length scale). However, a flow that is characterized by a small Rossby number does not, by itself, imply that inertia is not important in the dynamics. It is well known that small Rossby number flows can be highly turbulent, i.e.~that the available potential energy in small departures from geostrophic balance is substantial \citep[e.g.][]{jP87}. It is therefore necessary to rely on employing either the large Prandtl number limit, or to arbitrarily restrict the flow to weakly supercritical states in order to justify rigorously the neglect of inertia; both of these approximations require the Reynolds number $Re = U L/\nu$ to be small.


Despite the significant numerical advantages associated with eliminating inertia, the resulting set of equations still possess significant stiffness owing to the intrinsic separation in scales that occurs when the Rossby number is small. The mathematical result of this limit is that spatial derivatives perpendicular to the rotation axis become asymptotically larger than derivatives parallel to the rotation axis \citep{sC61} and that fast inertial waves become weakly damped. For this reason, the computational models that have employed the large Prandtl number approximation are still unable to reach the relevant geophysical limit of small Ekman number, $Ek = \nu / 2\Omega L^2 \ll 1$. It is therefore advantageous to pursue further reduction strategies that exploit the scale separation of rapidly rotating convection with the use of multiscale asymptotics \citep[e.g.][]{kJ07}. Such an approach has proven invaluable for investigating the onset of linear convection in spherical geometries \citep{pR68,fB70,cJ00,eD04}, weakly nonlinear dynamo action in the plane layer geometry \citep{sC72,aS74,yF82}, and for the development of a fully nonlinear reduced convection model in the plane layer \citep[][]{mS06,kJ12}. \cite{mC15b} recently extended the Childress-Soward weakly nonlinear dynamo model by developing a fully nonlinear reduced dynamo model that is capable of simulating dynamo action well above the onset of convection; we refer to this new reduced model as the quasi-geostrophic dynamo model (QGDM). In the present work we extend the QGDM to the large Prandtl number limit for the purpose of investigating the physical ramifications.

\section{Model Development}

The development given here closely follows that of \cite{mC15b}. Since our interest is on the small-scale convective dynamics of planetary interiors, we consider the simplified rotating plane layer geometry in which a fluid layer of depth $H$ is confined between two horizontal boundaries. Such a geometry might be considered a simplified version of the region just above and below the inner core, known as the tangent cylinder \citep[e.g.][]{jmA03}. The system rotates about the vertical axis with rotation vector $\rot = \Omega \hz$, and the gravity vector is $\mathbf{g} = - g \hz$. We utilize the Oberbeck-Boussinesq approximation and denote the fluid density as $\rho$. The fluid layer is heated from below and cooled from above by a constant temperature difference $\Delta \Theta$. The governing equations are written as
\begin{gather}
\dst \ub + \ub \cdot \nabla \ub + \frac{Pr}{Ek} \, \hz \times \ub = - \frac{Pr^2}{Ek^2} \, \nabla p + M \, \Bb \cdot \nabla \Bb + Ra Pr \, \vartheta \, \hz + Pr \nabla^2 \ub, \label{E:mom1} \\
\dst \vartheta + \div \lb \ub \vartheta \rb =  \nabla^2 \vartheta, \\
\dst \Bb  =  \nabla \times ( \ub \times \Bb) +  \frac{1}{q} \nabla^2 \Bb , \\
\div \ub =  0 , \quad \div \Bb  =  0 \label{E:Sol1} ,
\end{gather}
where we have non-dimensionalized utilizing the small horizontal scale of the convection $\ell$, compositional diffusion time $\ell^2/D$, modified pressure $\rho (2\Omega \ell)^2$, magnetic field magnitude $\mathcal{B}$ and temperature $\Delta \Theta$. To simplify notation, the subscript on the Prandtl number is omitted with the assumption that we are considering a compositional Prandtl number in the context of planetary interiors. The Rayleigh number and the compositional Roberts number are defined by 
\be
Ra =\frac{g \gamma \Delta \Theta \ell^3}{\nu D },\quad 
q = \frac{D}{\eta} ,
\label{E:params1}
\ee
where the thermal expansion coefficient is $\gamma$. The small-scale Ekman and Rayleigh numbers are related to their large-scale counterparts by the relations
\be
Ek = Ek_H \lb \frac{H}{\ell} \rb^2, \quad Ra = Ra_H \lb \frac{\ell}{H} \rb^3 . 
\ee

In its most general form, the parameter $M$, which is the ratio of magnetic to kinetic energy densities, is given by
\be
M = \frac{\mathcal{B}^2}{\rho \mu \, \mathcal{U}^2} = \lb \frac{\mathcal{U}_\mathcal{A}}{\mathcal{U}} \rb^2,
\ee
where the Alfv\'en velocity is defined as $\mathcal{U}_\mathcal{A} = \mathcal{B}/\sqrt{\rho \mu}$ and $\mu$ is the magnetic permeability. Thus, $M$ is also the square of the inverse Alfv\'en number. With the compositional diffusion scaling $\mU = D/\ell$, $M$ becomes 
\be
M = \frac{\mathcal{B}^2 \ell^2}{\rho \mu \, D^2}. \label{E:Mdiff}
\ee

In the limit of rapid rotation it is well known that convection becomes highly anisotropic with aspect ratio $H/\ell = Ek^{-1} \gg 1$, where $H$ is the depth of the fluid layer \citep{sC61}. This inherent scale separation is exploited with the use of multiscale asymptotics such that the axial space derivative and time derivative are written as \citep[cf.][]{mC15b}
\be
\dsz \rightarrow \dsz + Ek \dz , \quad
\dst \rightarrow \dst + A_\tau^{-1} \dtau + Ek^2 \dt .
\ee
The slow axial coordinate is $Z = Ek \, z$, the convective timescale is given by $t$, the slow mean magnetic field timescale is $\tau = A_\tau^{-1} t$ and the slow mean temperature timescale is $T = Ek^2 t$. We find that the precise definition of $A_\tau$ depends upon the asymptotic ordering of $q$ employed, and so will be given explicitly in later sections. The Proudman-Taylor constraint of vertical invariance is satisfied on the small axial scale $z$; to allow for convection on the large-scale axial coordinate $Z$ this constraint must be broken at $O(Ek)$.

Each dependent variable is decomposed into mean and fluctuating components. For example, the generic variable $f$ is written as
\be
f (\litx,t, Z,\tau,T) = \overline{f} (Z,\tau,T) + f' (\litx,t, Z,\tau,T), \label{E:decomp}
\ee
where 
\be
\overline{f}(Z,\tau,T)  = \lim_{t', \mathcal{V} \rightarrow \infty} \, \frac{1}{t' \mathcal{V}} \int_{t',\mathcal{V}} f(\litx,Z,t,\tau,T) \, d \mathbf{x} \, dt , \quad \overline{f'} \equiv 0 ,
\ee
and $\mathcal{V}$ is the small-scale fluid volume. The fast and slow coordinates are given by $(\litx,t)$ and $(Z,\tau,T)$ respectively. To simplify the presentation we focus only on the case when no large-scale horizontal modulation is present. The result of this simplification is that the mean velocity and the mean axial magnetic field are both zero; use of this fact is made from the outset.


Utilizing decompositions of the form \eqref{E:decomp} for each variable, the equations are separated into mean and fluctuating components. Averaging the equations over fast temporal and spatial scales results in the mean equations
\begin{gather}
Ek \dz \overline{\lb w' \ubp \rb} =   - \frac{Pr^2}{Ek} \dz \mP \, \hz + M \overline{\mathbf{F}}_L + Ra Pr \, \overline{\theta} \, \hz , 
\label{E:meanmom} \\
Ek^2 \dt \mth +  Ek \dz \overline{\lb \pth w'  \rb}  =  Ek^2 \dzt \mth \label{E:meantemp} , \\
\oAt \dtau \mBb =  Ek \hz \times \dz \overline{ (\ubp \times \bbp)}  +  \frac{Ek^2}{q} \dzt \mBb \label{E:minduc0},
\end{gather}
where the mean Lorentz force is $\overline{\mathbf{F}}_L = \overline{\bbp \cdot \nabla \bbp}$ and Gauss's law is trivially satisfied for the mean magnetic field in the absence of large-scale horizontal modulations. 

The fluctuating equations are obtained by subtracting the mean equations from the complete equations to yield
\begineq
\begin{split}
\lb \dst + \oAt \dta + Ek^2 \dt \rb \ubp  + \nabla \cdot ( \mathbf{u} \mathbf{u}^{\textnormal{T}} ) + Ek \dz( w' \ubp )   - Ek \dz \overline{\lb w' \ubp \rb}  + \\ \frac{Pr}{Ek} \, \hz \times \ubp =  - \frac{Pr^2}{Ek^2} \, \lb \nabla +  \hz Ek \dz  \rb \pP + M \, \mathbf{F}_L^{\prime}+ Ra Pr \, \thp \, \hz + \\ Pr \lb \nabla^2 + Ek^2 \dzt \rb \ubp  ,
\label{E:flucmom}
\end{split}
\endeq
\begineq
\begin{split}
\lb \dst + \oAt \dta + Ek^2 \dt \rb \pth + \div \lb  \ubp \pth \rb + \\  Ek \dz \lb \ubp \pth \rb  +  Ek w' \dz \mth  - Ek \dz \overline{ \lb \ubp \pth \rb}    =   \lb  \nabla^2 + Ek^2 \dzt  \rb \pth ,
\end{split}
\endeq
\begineq
\begin{split}
\lb \dst + \oAt \dta + Ek^2 \dt \rb \bbp  = \\  Ek \hz \times \dz ( \ubp \times \mBb) +   Ek \hz \times \dz ( \ubp \times \bbp) - Ek \hz \times \dz \overline{(\ubp \times \bbp)} + \\ \ +   \nabla \times ( \ubp \times \mBb)  +  \nabla \times ( \ubp \times \bbp)  + \frac{1}{q} \lb \nabla^2 + Ek^2 \dzt \rb \bbp \label{E:finduc0} ,
\end{split}
\endeq
\begineq
\lb \nabla + Ek \dz \rb \cdot \ubp = 0, \quad \lb \nabla + Ek \dz \rb \cdot \bbp = 0 , \label{E:flucsolen}
\endeq
where the superscript ``\text{T}" denotes a transpose.



The mean and fluctuating equations are reduced by representing each dependent variable with a double asymptotic expansion in powers of $Ek^{1/2}$ and $Pr^{-1}$, and taking the limits $(Ek,Pr^{-1}) \rightarrow 0$. For example, the fluctuating velocity field becomes
\be
\ubp = \sum_{i,j} Ek^{i/2} Pr^{-j} \ubp_{i/2,j} .
\ee

The derivation of the reduced (mean and fluctuating) induction and heat equations is identical to those given previously in \cite{mC15b}. To avoid repetition we therefore focus only on the differences between \cite{mC15b} and the present work.

The mean momentum equation gives hydrostatic balance at $O(Pr/Ek)$
\be
\dz \mP_{0,1} = \Rat \, \mth_{0,0},
\ee
where the reduced Rayleigh number is defined as 
\be
\Rat = Ra Ek = Ra_H Ek_H^{4/3} = O(1).
\ee

Geostrophy appears for several orders in the fluctuating momentum equation, with the leading order expression at $O(Pr Ek^{-1})$ 
\be
\hz \times \ubp_{0,0} = - \nabla \pP_{1,1}. 
\ee
Similarly, the continuity equation gives
\be
\div \ubp_{0,0} = 0.
\ee
We can then define the geostrophic streamfunction as $\psi'_{0,0} \equiv \pP_{1,1}$ such that $\ubp_{0,0} = (-\dsy \psi'_{0,0}, \dsx \psi'_{0,0}, w'_{0,0})$. 


At $O(Pr)$ the momentum equation is 
\be
 \hz \times \ubp_{1,0} = - \nabla \pP_{2,1} - \dz \psi'_{0,0} \hz + \Rat \pth_{1,0} \hz + M \mathbf{F}_L^{\prime} + \lp \ubp_{0,0} . \label{E:fmom}
\ee
In agreement with previous work \citep{mC15b}, we require that $\pth = O(Ek) =  \pth_{1,0}$.

Solvability consists of operating on equation \eqref{E:fmom} with $\hz \cdot \nabla \times$ and $\hz \cdot$ and averaging over the small axial scale $z$ to obtain the fluctuating vorticity and axial momentum equations
\be
 - \dz \pfw_{0,0} =  M \, \hz \cdot \nabla \times \mathbf{F}_L^{\prime}  +  \nabla_\perp^4 \psi_{0,0} , 
\ee
\be
 \dz \psi_{0,0} = \Rat \pth_{1,0} +  M \, \hz \cdot \mathbf{F}_L^{\prime}  +  \lp \pfw_{0,0} ,
\ee
where use has been made of the continuity equation at $O(Ek)$
\be
\div \ubp_{1,0} + \dz w'_{0,0} = 0 .
\ee
Apart from the Lorentz force, these equations are identical to those studied in \cite{mS06}.

The reduced mean and fluctuating heat equations are given by
\be
\dt \mth_{0,0} + \dz \overline{\lb w'_{0,0} \pth_{1,0}\rb} = \dzt \mth_{0,0} ,
\ee
\be
\dst \pth_{1,0} + \ubp_{0,0} \cdot \nabla_\perp \pth_{1,0} + w'_{0,0} \dz \mth_{0,0} = \lp \pth_{1,0} .
\ee

At this point the Lorentz force has been kept in general form. For a saturated dynamo state to be possible, the Lorentz force must enter at the same asymptotic order as the other terms present in equation \eqref{E:fmom} such that
\be
M |\mBb| |\bbp| = O(Pr).
\ee
If we take $|\mBb| = O(1)$, as understood by the definition of $M$, we find two possibilities that both hinge on $|\bbp| = O(q)$ to allow magnetic diffusion to be important on the small horizontal scales; these are $q = O(1)$ and $q=O(Ek^{1/2})$. We discuss each of these limits in the following two subsections.

\subsection{The $q\ll 1$ limit}

The geophysically relevant, low $q$, regime is characterized by the distinguished limit $q=Ek^{1/2}\wq$, where $\wq$ is an order one, asymptotically reduced, compositional Roberts number. For this limit the magnetic to kinetic energy density ratio becomes
\be
M = O(Pr Ek^{-1/2}),
\ee
and the fluctuating Lorentz force simplifies to
\be
\mathbf{F}_L^{\prime} = \mBb_{0,0} \cdot \nabla_\perp \bbp_{1/2,0} .
\ee
The reduced induction equations are given by
\begin{gather}
\dtau \mBb_{0,0} = \hz \times \dz \overline{ (\ubp_{0,0} \times \bbp_{1/2,0})} +  \frac{1}{\qt} \dzt \mBb , \label{E:mmag1} \\
0 =   \mBb_{0,0} \cdot \nabla_\perp \ubp_{0,0} + \frac{1}{\qt} \lp \bbp_{1/2,0} ,\label{E:fmag1} 
\end{gather}
where the mean magnetic field timescale is $\tau = Ek^{3/2} t$ such that $A_\tau = Ek^{-3/2}$.

Thus, we find that for dynamo action to occur in this limit the fluctuating magnetic field must be asymptotically weaker than the mean magnetic field and only ``mean-eddy" interactions are important in the Lorentz force. The reduced mean magnetic field equation \eqref{E:mmag1} shows that temporal changes arise from imbalances in ohmic diffusion on the large vertical scale $Z$ and the electromotive force (emf). In contrast, the reduced fluctuating magnetic energy equation is the well-known quasi-static form whereby the fluctuating magnetic field adjusts instantaneously relative to the fluctuating velocity field.

\subsection{The $q=O(1)$ limit}

The $q=O(1)$ limit is representative of the parameter regime accessible by direct numerical simulation (DNS) \citep[e.g.][]{sS04,dH16}. For this limit the mean and fluctuating magnetic fields are of the same asymptotic order and the magnetic to kinetic energy density ratio scales as
\be
M = O(Pr).
\ee
The reduced Lorentz force becomes 
\be
\mathbf{F}_L^{\prime} = \mBb_{0,0} \cdot \nabla_\perp \bbp_{0,0} + \bbp_{0,0} \cdot \nabla_\perp \bbp_{0,0}.
\ee
In contrast to the low $q$ limit, both mean-eddy and eddy-eddy interactions are important in the saturation of the magnetic field. The reduced induction equations are given by
\begin{gather}
\dtau \mBb_{0,0} = \hz \times \dz \overline{ (\ubp_{0,0} \times \bbp_{0,0})} , \label{E:mmag2} \\
\dst \bbp_{0,0} + \ubp_{0,0} \cdot \nabla_\perp \bbp_{0,0}  = \lb \mBb_{0,0} + \bbp_{0,0} \rb \cdot \nabla_\perp \ubp_{0,0} + \frac{1}{q} \lp \bbp_{0,0} , \label{E:fmag2}
\end{gather}
where the mean magnetic field evolves on a faster timescale $\tau = Ek \, t$ relative to the low $q$ limit.

Here we see from equation \eqref{E:mmag2} that ohmic diffusion is subdominant relative to induction on the large vertical scale. The fluctuating induction equation \label{E:fmag2} shows that horizontal advection of the fluctuating magnetic field becomes important, and the first-order-smoothing approximation is no longer valid in this limit \citep[e.g.~see][]{hM78b}.

\section{Discussion}

In the present work we have developed two self-consistent asymptotic dynamo models valid for inertia-less, large Prandtl number, rapidly rotating convection. In contrast to the previous work of \cite{mC15b}, it is the Roberts number $q$, rather than $Pm$, that becomes the important diffusivity parameter in the final model. However, we can relate these two parameters upon noting that $q = Pm/Pr$. We emphasize here that $q$ should be interpreted in terms of a compositional Roberts number. For $q = O(1)$, this implies that $Pm = O(Pr)$, i.e.~$Pm \gg 1$, which is clearly not relevant geophysically. For the $q = O(Ek^{1/2})$ model we have $Pm = O(Pr Ek^{1/2})$; to obtain $Pm \ll 1$ we require that $Pr < Ek^{-1/2}$, or in terms of $Ek_H$ this becomes $Pr < Ek_H^{-1/6}$.  This latter requirement is fully consistent with the asymptotics, showing that it is possible to drive a low $Pm$, low $q$ dynamo with inertia-less convection in the limit of rapid rotation. Recalling that for the Earth's core $Ek_H = O(10^{-15})$, so that the large Prandtl number approximation is only valid provided $Pr \lesssim O(300)$; studies suggest that this inequality does hold for most chemical species \citep{mP13}.

We note that both the small $q$ model of the present work and the small $Pm$ model of \cite{mC15b} are small magnetic Reynolds number models, $Rm = U\ell/\eta \ll 1$. In addition, the small $q$ model is characterized by a small convective-scale Reynolds number, $Re = U\ell/\nu \ll 1$; this follows from the fact that the P\'eclet number $Pe = U\ell/D = Pr Re = O(1)$ to allow convective heat transport. How, then, are such flows capable of driving small $Pm$ dynamos in light of the relation $Rm = Re Pm$? The answer lies in the fact that the large-scale magnetic Reynolds number, $Rm_H = Rm/Ek$, remains large since we have
\be
Rm_H = \frac{Rm}{Ek} = Ek_H^{-1/6} \gg 1 .
\ee
Thus, magnetic induction dominates ohmic diffusion on the large ($Z$) scale, consistent with the fact that both of the models presented here, and those developed in \cite{mC15b}, are large-scale dynamos. Indeed, setting the mean field to be identically zero in any of the variations of the QGDM will eliminate all dynamo action. 

%


We can also place a bound on the Reynolds number by utilizing the relation $Rm = Re Pm$. For the small $Rm$ limit this implies that $Re > Ek^{1/2}$ for the fluid to be considered low $Pm$ since $Re = Pe/Pr = O(1/Pr)$. The neglect of inertia is therefore limited to a finite range in $\Rat$ since $Re$ will grow as the buoyancy forcing is increased. Since $Re$ is an unknown function of $\Rat$, we require detailed numerical simulations of the QGDM to determine the appropriate range in $\Rat$ over which the large Prandtl number limit remains accurate.

Table \ref{T:lims} shows the various asymptotic limits employed in the present work and the $Pr=O(1)$ cases of \cite{mC15b}. Taken together, these investigations show that the ratio of magnetic to kinetic energy densities, $M$, is not indicative of dominant balances in the governing equations since all of these models are geostrophically balanced. Rather, the size of $M$ is related to the presence of Alfv\'en waves in the reduced dynamics; order one values of $M$ indicate the presence of these waves, whereas such waves are filtered if $M \gg 1$. This filtering can be obtained by asymptotically eliminating the time derivative of either the velocity field in the fluctuating momentum equation or the time derivative of the magnetic field in the fluctuating induction equation. As shown in the present work, the elimination of both of these time derivatives yields a separation in the convective and Alfv\'en timescales that is larger than when only eliminating one of these time derivatives. 

Consistent with DNS, only for the $(Pr,Pm)=O(1)$ case do we find that $M=O(1)$. There is very little available data for small $q$ or small $Pm$ due to the computational cost of such studies, so no comparison can be made at present with the asymptotic relations for these limits. However, \cite{sS04} have reported values of $M$ (denoted as $E_{mag}/E_{kin}$ in their Table II) for all of their DNS cases with $q=1$. Of particular interest for comparison with the current work are their $Ek_H = 5 \times 10^{-6}$ cases for which they consider the three Prandtl numbers $Pr=1$, $10$, and $30$, where the Rayleigh number is the same for all three cases. In order of increasing Prandtl number they find that $M \approx 2, 11$, and $28$, respectively. These computed values compare well with the predictions $M=O(1)$, $O(10)$, and $O(30)$ based on the $Pr=O(1)$ model of \cite{mC15b} and the present $q=O(1)$ model.

\begin{table}
  \begin{center}
    \begin{tabular}{lcccc}
      $Pr$      &     $q$  &   $Pm$  &  $M$    \\
      \hline
      \hline
      $\gg 1$       &  $O(1)$   &   $O(Pr)$   &  $O(Pr)$     \\   
      $\gg 1$       &  $O(Ek^{1/2})$  &     $O(Pr Ek^{1/2})$  &     $O(Pr Ek^{-1/2})$     \\      
      $O(1)$       &  $O(1)$   &     $O(1)$  &     $O(1)$   \\   
      $O(1)$       &  $O(\ep^{1/2})$  &  $O(\ep^{1/2})$  &     $O(\ep^{-1/2})$     \\
\end{tabular}
    \caption{Various asymptotic limits for the dynamo models developed in the present work ($Pr \gg 1$) and in \cite{mC15b} where $Pr=O(1)$. $Pr$ is the Prandtl number, $q$ is the Roberts number and $Pm = q Pr$ is the magnetic Prandtl number. The parameter $M$ is the ratio of magnetic to kinetic energy densities. The Rossby number is given by $\ep = Re Ek$, where $Re$ and $Ek$ are the Reynolds and Ekman numbers based on the small convective scale.}
    \label{T:lims}
  \end{center}
\end{table}

Although the present models are significantly oversimplified with respect to the Earth's outer core, it is nevertheless a useful exercise to determine if the asymptotic limits taken here are consistent with what is known about the geodynamo. Studies suggest a magnetic field strength in the core of $\mathcal{B} \sim 1 \, \text{mT}$ \citep[e.g.][]{nG10} and a flow speed of $\mathcal{U} \sim 10^{-4} \, \textnormal{m} \, \textnormal{s}^{-1}$ \citep[e.g.][]{cF11}, leading to an Alfv\'en speed of $\mathcal{U}_\mathcal{A} \sim 10^{-2} \, \textnormal{m} \, \textnormal{s}^{-1}$. These values lead to an estimate of $M \sim 10^4$ where, to remain consistent with the asymptotic derivation, we assume that the observed velocity scales diffusively as given by equation \eqref{E:Mdiff}. Since we know that $(q,Pm) \ll 1$ in the outer core, the small $q$ inertia-less model of the present work and the small $Pm$ model of \cite{mC15b} give values of $M \sim 3\times10^4$ and $M \sim 3\times10^2$, respectively, where we have assumed that $Ek_H = 10^{-15}$ and $Pr=100$; utilizing smaller values of $Pr$ will yield corresponding reductions in our estimate of $M$. Although the large $Pr$ asymptotic estimate of $M$ is in better agreement with our observational estimate of $M$, we emphasize that there is enough variation and uncertainty in the values of $\mathcal{U}$ and $\mathcal{U}_\mathcal{A}$ that may imply the $Pr=O(1)$ model is more relevant to core dynamics. Nevertheless, such a large value of $M$ for the core does suggest that Alfv\'en waves are not likely to be important for the geodynamo on the small scales of convection, though they can play an important role on the large scales of the core \citep[e.g.][]{nG10}. Simulations of the reduced dynamo models will better allow us to assess which approximation is more suitable for understanding the geodynamo. Recent kinematic investigations for the $Pr=O(1)$ QGDM are reported in \cite{mC16} and \cite{mC16b}.

It is common in the literature to distinguish dynamo models based on the strength of the self-generated magnetic field. \cite{sC72} defined the field strength based on the asymptotic ordering of the Hartmann number $Ha_H = \mathcal{B} H/ \lb \rho \mu \eta \nu \rb^{1/2} = Ek_H^{-1/6} \lb q M / Pr \rb^{1/2}$, and the deviation of the Rayleigh number from its critical value. With regards to the \cite{sC72} terminology, the weak-field and intermediate-field regimes were investigated by \cite{aS74} and \cite{yF82}, and characterized by Hartmann numbers $Ha_H = O(1)$ and $Ha_H = O(Ek_H^{-1/6})$, respectively. All four of the QGDM variations identified thus far (listed in Table \ref{T:lims}) are characterized by a Hartmann number $Ha_H = O(Ek_H^{-1/3})$. However, we emphasize that both \cite{aS74} and \cite{yF82} were weakly nonlinear investigations in the sense that the mean temperature profile is always near that of the conductive state. As a result, the ratio of magnetic to kinetic energy $M$ is asymptotically small for both \cite{aS74} and \cite{yF82} \citep[e.g.~see][]{mC15b}. Therefore, both weak-field and intermediate-field dynamos have a magnetic energy that is asymptotically smaller than the kinetic energy in the flow. This characteristic is in stark contrast to the four variations of the QGDM, which describe \textit{saturated} dynamos that can have magnetic energy of the same order as, or asymptotically larger than, the kinetic energy. It is thus an oversimplification to categorize turbulent dynamos in rapidly rotating systems as either weak- or strong-field. Indeed, the Lorentz can have just as much influence on the quasi-geostrophic flow as does the buoyancy force in the QGDM, as demonstrated by equation \eqref{E:fmom}.

An important question is whether there is a signature of the predominant forcing mechanism for the geodynamo in geomagnetic field observations. For instance, does the morphology of the geomagnetic field, or behavior of the secular variation depend upon whether thermal or compositional forcing is dominant in the core? Convection simulations find that the dynamics can indeed exhibit quite distinct behavior depending upon the value of the Prandtl number \citep[e.g.][]{mB10,mC12b}. The dynamics become further complicated when both forcing mechanisms are present \citep{tT12,fT14}. Dynamo simulations find that this Prandtl number dependence can result in significant changes in the structure of the resulting magnetic field \citep{fB06,bS06}. Moreover, `two-and-a-half' dimensional models find that inertia changes the temporal character the resulting dynamo \citep{dF01}, and mean-field models have found that the presence of inertia tends to facilitate dynamo action and therefore leads to stronger magnetic fields \citep{dF04}. The present model will allow for an extension of this previous work to the limit of rapid rotation and realistic fluid properties.


\section*{Acknowledgments}
This work was supported by the National Science Foundation under award number EAR-1620649 (MAC and KJ).

\bibliography{../Dynamobib}

\end{document}